# A steady-state stability analysis of uniform synchronous power grid toplologies

*James R. Stright[a],\*, Chris S. Edrington[a]*

[a] *Department of Electrical and Computer Engineering, FAMU-FSU, Tallahassee 32301, USA*

ABSTRACT

Motter et al. derived a real-valued master stability function which determines whether and to what degree a given power grid is asymptotically stable. Stright and Edrington adopted certain uniformity assumptions on a grid's components and demonstrated how differences in topologies obtained using these components can affect the stabilities of the resulting grids. Building on this work, we show via simulations the physical significance of stability as opposed to instability. We show that for stable topologies, increased stability can correspond to decreased generator torque ripple. We also describe how some elementary changes in grid topology can affect stability values. Known stability values for certain abstract circulant grids are used to quantify stability enhancement as interconnection density increases.

## 1. Introduction

The steady-state stability of electric grids powered by synchronous generators was examined by Motter et al. [1] and subsequently by Stright and Edrington [2]. This work is briefly reviewed and then extended to include some interpretations and conjectures based on computer simulations.

Graph theory is widely used to model electric power grids. An elementary but useful approach represents a grid as a connected graph with two kinds of nodes, synchronous generators and substations (or loads), and with edges corresponding to transmission or distribution lines [1]. As dynamic entities, grids are constantly subjected to disturbances to their steady-state performance, which can be regarded as maintaining a nearly constant grid frequency. These disturbances range from the normal addition and removal of various kinds of loads to malfunctions resulting from, for example, severe weather conditions. Disturbances typically affect the rotational velocities of one or more generators, but these generators usually return quickly to their pre-disturbance rotational speeds. This is the defining characteristic of steady-state stability as it applies to power grids; it is what is meant by the ability of a grid to synchronize spontaneously.

---

\* *Corresponding author.*
E-mail address: js14w@my.fsu.edu (James R. Stright)

Edges of graphs which join generators will be shown particularly important in steady-state analyses of power grids. With appropriate uniformity assumptions, these edges form the "skeletons" of grids which survive the algebraic process of Kron reduction, a process that essentially condenses the dynamical interactions of generators and loads to interactions among "virtual" generators equal in number to the number of generators in the underlying grid.

There is considerable diversity among components of all real-world grids. Synchronous generators vary widely in size, and loads and transmission lines vary widely in their impedance characteristics. Because of this diversity, it is difficult to gain fundamental understanding of the impact on steady-state stability of reconfigurations of a given set of grid components, say by replacing a transmission line between two generators with a transmission line joining two others. However, by adopting certain uniformity assumptions on grid components, it becomes possible to compare directly the relative stabilities of different grid configurations, i.e., of different grid topologies.

An appropriate set of such assumptions will be presented, including assumptions on the nature and placement of loads. These assumptions will permit graphical representations of abstract grids consisting only of "virtual" generators. Different choices of edges joining these generators then determine different topologies, and the relative stabilities of these topologies will be compared. The uniformity assumptions, while not applicable to any extant power grid, nevertheless provide a baseline for an understanding of how real grid topologies might influence grid steady-state stability.

Special attention will be given to the least densely connected topologies, that is, to topologies represented by tree graphs. The influence of topological density on steady-state stability will be examined via grids represented as regular circulant graphs. These grids can be parameterized to grow from sparsely to densely connected, and explicit stability values can be determined for all of them.

## 2. Background

Motter et al. derive a necessary and sufficient condition for a grid to be steady-state stable by first representing the grid as an admittance matrix [1]. If the grid has $n$ generators and $m$ loads with the $n$ generators indexed first, this is an $(n+m) \times (n+m)$ matrix they denote $\boldsymbol{Y_0}$. Its entries are linear combinations of various admittances associated with the grid [3]. Kron reduction then replaces this matrix with an $n \times n$ matrix $\boldsymbol{Y}$ whose complex-valued off-diagonal elements $Y_{rs} = G_{rs} + jB_{rs}$ can be regarded as encapsulating the essence of the grid's dynamics via connections among the grid's $n$ generators. For all $i = 1, \dots, n$, the internal voltage magnitude of the $i^{th}$ generator is denoted $E_i$, its inertia constant $H_i$, its steady-state rotational phase $\delta_i^*$, and its steady-state angular velocity $\omega_i^*$. The grid's reference frequency is denoted $\omega_R$. They assume the $n$ generators are operating at steady-state pairs of values $(\delta_i^*, \omega_i^*)$. They regard these $n$ pairs as an equilibrium point in $\mathbb{R}^{2n}$ and linearize the swing equation about that point, thereby obtaining an $n \times n$ coupling matrix denoted $\boldsymbol{P}$. The $(i,j)^{th}$ element of $\boldsymbol{P}$ is given by

$$P_{ij} = \begin{cases} \dfrac{\omega_R E_i E_j}{2H_i}\left(G_{ij}\sin\delta^*_{ij} - B_{ij}\cos\delta^*_{ij}\right), & i \neq j \\ -\displaystyle\sum_{k\neq i} P_{ik}, & i = j \end{cases} \quad (1)$$

where $\delta^*_{ij} = \delta^*_i - \delta^*_j$.

Since each diagonal element $P_{ii}$ of $\boldsymbol{P}$ is the negative of the sum of the off-diagonal elements in the same row, $\boldsymbol{P}$ has an eigenvector of all ones corresponding to a zero eigenvalue. However, because the underlying grid is assumed connected, $\boldsymbol{P}$ has only one zero eigenvalue [4]. A uniformity assumption that will henceforth be applied is that all generators are identical. In particular, all generators have the same rotational phase so in (1) all $\delta^*_{ij} = 0$. Indeed, with this uniformity assumption, $\boldsymbol{P}$ is a real symmetric matrix and therefore has $n$ real eigenvalues. These are denoted $\alpha_1, \alpha_2, \ldots, \alpha_n$ with $\alpha_1$ being the zero eigenvalue and the remaining eigenvalues arranged in nondecreasing order $\alpha_2 \leq \alpha_3 \leq \cdots \leq \alpha_n$. The smallest nonzero eigenvalue is thus $\alpha_2$, which Motter et al. refer to as the grid's stability value. They demonstrate that $\alpha_2$ determines whether the grid is steady-state stable. If $\alpha_2 < 0$ the grid is unstable, but if $\alpha_2 > 0$ the grid is stable.

The next section will derive the coupling matrix $\boldsymbol{P}$ from an admittance matrix $\boldsymbol{Y_0}$ corresponding to a small grid. It will be seen that the shunt admittances associated with each of the two generators represented in $\boldsymbol{Y_0}$ are not represented in $\boldsymbol{P}$, meaning that shunt admittances are irrelevant to the grid's steady-state stability. This is true in general, as may be seen by the effect of Kron reduction on admittance matrices. Kron reduction is often described as an iterative process, basically eliminating one load at each step, but it can also be accomplished more succinctly via Schur complementation [5,6]. Suppose $Y_0$ is represented as

$$Y_0 = \begin{bmatrix} A & B \\ C & D \end{bmatrix} \tag{2}$$

where $A$ is an $n \times n$ matrix determined by the $n$ generators and $D$ is an $m \times m$ matrix determined by the $m$ loads. Schur complementation yields the Kron-reduced matrix

$$Y = A - BD^{-1}C. \tag{3}$$

Shunt admittances apply only to individual generators, so they do not appear at all in the matrix $BD^{-1}C$. They do appear in the diagonal entries of $A$, but not in its off-diagonal entries, so only the diagonal entries of $Y$ are affected by shunt admittances. But no diagonal entry of $Y$ is represented in (1), so shunt admittances have no influence on $P$.

Classical Kron reduction can actually be regarded as iterated Schur complementation [6]. Iterative calculation of the right side of (3), with $D$ a singleton at each step, will also yield the matrix $Y$.

Regardless of how **Y** and then **P** are obtained, with the stated assumptions **P** is a real symmetric matrix. So, too, are the Laplacian matrices associated with all undirected simple graphs [4]. The smallest nonzero eigenvalue of a graph's Laplacian matrix is known as the graph's algebraic connectivity [7]. Algebraic connectivities of graphs have been studied extensively in several engineering contexts [8,9].

**3. An example grid**

The process of determining a stability value $\alpha_2$ will be illustrated for a small grid whose admittance diagram is depicted in Fig. 1. Nodes 1 and 2 represent identical generators and node 3 is a load. The conductances of the depicted admittances $Y_{0a}$, $Y_{0b}$, and $Y_{rs}$ will be considered negligible and the negatives of their respective susceptances will be denoted $k_a$, $k_b$, and $k_{rs}$. The admittance matrix corresponding to Fig. 1 can then be written

$$\begin{bmatrix} -(k_a + k_{12} + k_{13})j & k_{12}j & k_{13}j \\ k_{12}j & -(k_b + k_{12} + k_{23})j & k_{23}j \\ k_{13}j & k_{23}j & -(k_{13} + k_{23})j \end{bmatrix} \quad (4)$$

where $j = \sqrt{-1}$.

If $k_{13} \neq -k_{23}$, the Kron-reduced form of this matrix is

$$\mathbf{Y} = j \begin{bmatrix} -k_a - k_{12} - k_{13} + \frac{k_{13}^2}{k_{13}+k_{23}} & k_{12} + \frac{k_{13}k_{23}}{k_{13}+k_{23}} \\ k_{12} + \frac{k_{13}k_{23}}{k_{13}+k_{23}} & -k_b - k_{12} - k_{23} + \frac{k_{23}^2}{k_{13}+k_{23}} \end{bmatrix}. \quad (5)$$

The off-diagonal entries of this symmetric matrix $\mathbf{Y}$ are

$$Y_{12} = Y_{21} = G_{12} + jB_{12} = 0 + j\left(k_{12} + \frac{k_{13}k_{23}}{k_{13}+k_{23}}\right) \quad (6)$$

so $B_{12} = k_{12} + k_{13}k_{23}/(k_{13} + k_{23})$.

With both generators identical, the term $\omega_R E_i E_j / 2H_i$ in (1) is a constant that will be assumed one. This allows $\mathbf{P}$ to be written

$$\mathbf{P} = \begin{bmatrix} P_{11} & P_{12} \\ P_{21} & P_{22} \end{bmatrix} = \begin{bmatrix} B_{12} & -B_{12} \\ -B_{12} & B_{12} \end{bmatrix} \tag{7}$$

which has eigenvalues

$$\alpha_2 = 2B_{12} = 2\left(k_{12} + \frac{k_{13}k_{23}}{k_{13}+k_{23}}\right). \tag{8}$$

Now suppose $k_{12} > 0, k_{23} > 0$, and $-k_{12}k_{23} < k_{13}(k_{12} + k_{23}) < 0$, where each $k_{rs}$ corresponds to an edge between nodes $r$ and $s$ in Fig. 1. With all edges present, and even with the edge between nodes 1 and 3 eliminated by setting $k_{13} = 0$, substitution into (8) reveals that $\alpha_2 > 0$, corresponding to stable topologies. However, setting only $k_{12} = 0$ gives $\alpha_2 < 0$, indicating an unstable topology. These grids' topologies and their corresponding stabilities are illustrated in Fig. 2.

These stability responses were obtained via simulation with MATLAB's SimPowerSystems, using a modified version of the file model6_1.slx provided in Viktor Perelmuter's book on electrotechnical systems [10]. Perelmuter presented this two-generator, one-load model to illustrate the use of the Simplified Synchronous Machine block provided by SimPowerSystems. This block allows convenient access to many synchronous generator parameters, including rotor angle velocity, and although it is not the most general synchronous generator model available in SimPowerSystems, it was adequate for the purposes of this research. Simulations with stop times of 13 seconds were performed for each topology in Fig. 2. A short-duration large resistive load (introduced for the interval from 3 to 3.1 seconds during the simulations) was added to the edge joining nodes 2 and 3 to compare the relative steady-state stabilities of the three topologies. With appropriate values for the parameters $k_{rs}$, damped sinusoidal responses after 3.1 seconds were observed for all rotor angle velocities except in the case of the unstable topology, as illustrated in Fig. 2.

The next section will reveal (as a special case) that if all three of $k_{12}$, $k_{23}$, and $k_{13}$ are equal and positive then every connected topology corresponding to Fig. 1 is stable.

## 4. Grids having identical admittances

Consider the case where all grid admittances (except possibly shunt admittances) have the same nonzero value $A$ and all shunt admittances have the same value $\varepsilon A$ where $\varepsilon > 0$. Let $v = n + m$ denote the total number of vertices in the grid (ignoring the reference node). As before, assume all $n$ generator nodes appear first in the node numbering scheme. Generalizing the notation of (4), each row $i$ of the symmetric $v \times v$ admittance matrix $Y_0$ has the form

$$-k_{i1}A \quad -k_{i2}A \quad \cdots \quad \left(\varepsilon_i + \sum_{\substack{l=1 \\ l \neq i}}^{v} k_{il}\right)A \triangleq k_{ii}A \quad \cdots \quad -k_{iv}A \qquad (9)$$

where $\varepsilon_i = \varepsilon$ if $1 \leq i \leq n$ but $\varepsilon_i = 0$ if $n + 1 \leq i \leq v$, and where each $k_{il}$, $l \neq i$, is either 1 (if an edge joins nodes $i$ and $l$) or 0 (if no edge joins nodes $i$ and $l$). Because the grid is assumed connected there is a path between every two nodes. In particular the $i^{th}$ node is connected to at least one other node so $Y_{0ii}$ is a positive multiple of $A$. By (9), this means that $Y_0$ is a diagonally dominant matrix [11]. Physical considerations permit the assumption that the matrix equation $I = Y_0 V$ has a unique solution, i.e., that $Y_0$ is invertible. It can be shown, making extensive use of Schur complementation, that Kron reduction preserves five properties of $Y_0$: its symmetry; its invertibility; its property of having nonpositive off-diagonal coefficients of $A$; its property of having diagonal entry coefficients of $A$ no less than the negative of the sum of all off-diagonal coefficients in the same row; and its property of having diagonal entries which are positive multiples of $A$ [2]. It follows that the Kron reduced matrix $Y$ is itself diagonally dominant.

With the additional assumption that $A$ represents a pure susceptance, $A = cj$ for some nonzero real number $c$, where $A$ denotes a pure inductance if $c < 0$ and a pure capacitance if $c > 0$. Together with the other assumptions presented in Section 3, this allows each $i^{th}$ row of the coupling matrix $P$ to be written

$$\begin{matrix} k_{i1}c & k_{i2}c & \cdots & -\sum_{\substack{l=1 \\ l \neq i}}^{n} k_{il}\, c = P_{ii} & \cdots & k_{in}c \end{matrix} \qquad (10)$$

where now all $k_{il} \geq 0$ with at least one $k_{il} > 0$. Thus $P$, too, is a diagonally dominant matrix. The Gersgorin disc theorem [11] can be invoked to show that if $c < 0$ then the underlying grid is stable but if $c > 0$ it is unstable.

This means that if all possible edges among grid components have the same purely inductive value, the relative stabilities of topologies formed by selective exclusion of edges can be directly compared. Some interesting examples of such topologies are considered next.

## 5. Analysis of stable topologies

As explained earlier, a grid's stability value is determined by an $n \times n$ coupling matrix $P$ which represents a "condensed" grid of $n$ "virtual" generators. Regardless of the number $m$ of loads in a grid, its stability value is derived from a "virtual" grid having only $n$ generators. To facilitate direct comparisons among grid topologies, a grid loading condition was derived which ensures that the loads in compared grid configurations have the same influence on grid stability, regardless of how the original grid generators are interconnected.

Of course, if a grid has no loads, its admittance matrix and its reduced admittance matrix are identical. Studying the block matrix form of an admittance matrix can reveal other ways in which certain grid loading structures can prevent the Kron reduction in (3) from reconfiguring an initial set of generator interconnections. Consider, for example, a grid in which each generator is connected to a load uniquely its own (and to no other load). Suppose all grid admittances have value $-j$. Assuming again that all $n$ generator nodes appear first in a grid's node numbering scheme, and now that the $n$ loads are numbered in the same order as their corresponding generators, the block form (2) of the grid's admittance matrix is

$$Y_0 = \begin{bmatrix} A & jI \\ jI & -jI \end{bmatrix} \qquad (11)$$

(where $A$ and $I$ are $n \times n$) and its reduced matrix has the form

$$Y = A - (jI)(-jI)^{-1}(jI) = A + jI. \qquad (12)$$

This type of grid loading therefore has no influence due to Kron reduction on the intergenerator edge weights represented by the matrix $A$ (i.e., on the off-diagonal elements of $A$). Therefore the coupling matrices for $Y$ and $A$ are identical, so the loads can be ignored for purposes of stability analysis. Furthermore, the values $G_{ij}$ and $B_{ij}$ in (1) are common to both $Y$ and $Y_0$.

Assuming any such loading condition, or any condition of unloaded or very lightly loaded generators, it becomes possible to depict grids and families of grids directly in Kron-reduced form, i.e., as connected generators only. This allows the use of a three-step process for mapping all possible connected grid topologies on a set of $n$ generators into positive real numbers $\alpha_2$, as illustrated in Fig. 3 for the case $n = 3$ and a complete topology. With these assumptions, the analysis of grid stability reduces to the analysis of connected graphs.

That being the case, a bit of terminology is in order [4]. A *graph* consists of a set of *nodes* and a set of *edges*, where an edge is an unordered pair of distinct nodes. Only graphs that have no loops or multiple edges (aka simple graphs) will be examined. If nodes $x$ and $y$ determine an edge, these nodes are said to be *adjacent*. A node is *incident* with an edge if it is one of the two nodes that define the edge. The *degree* of a node is the number of edges incident with it. If all nodes in a graph have the same degree, the graph is said to be *regular*. A *path* of length $l$ from node $x$ to node $y$ in a graph is a sequence of $l+1$ distinct nodes starting with $x$ and ending with $y$ such that consecutive nodes are adjacent. A graph is *connected* if there is a path between each pair of its nodes. Only connected graphs will be examined.

The physical meaning of the derived stability values depends on many factors, not least of which is the rather ambiguous notion of "small disturbance" in the definition of steady-state stability. One possible interpretation of the stability values $\alpha_2$ can be obtained by simulating a simple grid consisting of only two connected generators. Using SimPowerSystems and the same Simplified Synchronous Machine blocks described in conjunction with Fig. 2, but now with only the single edge between generators, simulations were performed using increasing magnitudes of inductive susceptance for that edge. Fig. 4 through Fig. 6 illustrate the changes in per-unit value of the rotor velocities of both generators as magnitudes of the inductive susceptances increased. Recalling that rotor angle acceleration is the derivative of rotor angle velocity, it can be seen that angle accelerations become smoother as stability values $\alpha_2$ increase. Rotor torque is directly proportional to rotor angle acceleration, so at least in this simple example, increasing stability values correspond to decreasing torque ripple. This is desirable not only for synchronous generators themselves but potentially for rotating machines throughout a grid [12].

Fig. 4 through Fig. 6 depict the effect on stability values of varying a connection's susceptance, but for purposes of comparing the relative stabilities of connected topologies all connections will be assumed to have identical inductive susceptances. In graph-theoretic terms, the most elementary connected topologies are trees, which in the present context connect all $n$ virtual generators with a minimal number $n-1$ of edges [4]. Some examples of trees having seven nodes are illustrated in Fig. 7.

The relative stabilities of grids corresponding to various types of trees were determined and compared. In graph theory, the *distance* between two nodes is the length of the shortest path between them, i.e., the minimum number of "hops" needed to reach one from the other. The *diameter* of a graph is the maximum distance between any two nodes. For example, the "star" graph in the middle of Fig. 7 has diameter 2, the smallest diameter of the three depicted graphs. Regarding the graphs of Fig. 7 as power grids with the assumptions previously discussed, the star grid also has the highest stability value; the grid on the left has the lowest. All experimentation performed to date suggests the conjecture that for any tree grid with a fixed number of nodes, grid stability increases as grid diameter decreases. This suggests a technique for finding a most stable way of joining any two tree grids with a single edge (thereby creating a single larger tree grid). For example, Fig. 8 illustrates that connecting the two central nodes of grids 1 and 2 yields a more stable grid than does connecting any other two nodes from these two grids. The addition of any other single edge between the two grids results in a tree having a larger diameter than the tree with connected central nodes (which has diameter 3), and true to the conjecture it also results in a tree having a lower stability value.

A corollary to this conjecture is that of all tree grids having three or more nodes, a star topology (with diameter two and stability value one) is the most stable, since every tree with three or more nodes necessarily has a diameter of at least two.

Tree grids have no redundant edges. Adding any edge to a tree necessarily creates a *cycle* within the resultant graph; that is, a closed path with distinct edges and nodes. The number of "hops" needed to follow the closed path from any initial node back to itself is the length of the cycle. All tree grids examined thus far suggest another conjecture, that in any tree grid with three or more nodes the addition of an edge which yields a cycle of longer length results in a more stable grid than does the addition of an edge which yields a cycle of shorter length. For example, the addition of the long, dotted edge in Fig. 9 (which creates a cycle of length 5) yields a grid with a higher stability value than the addition of the shorter dashed edge (which creates a cycle of length 3).

The relevance of several graph-theoretic properties to power grid structure has been extensively investigated by many researchers [13]. Properties other than diameter and cycle length include degree distribution, betweenness distribution, characteristic path length, and clustering coefficient. Certain families of graphs, much more richly interconnected than trees, may be particularly helpful in revealing the influence of such properties on grid steady-state stability. They can also reveal how stability can increase as a function of richness of interconnection. For example, families of *regular circulant graphs* can be constructed for any fixed odd integer $n \geq 3$ of nodes; such a family is illustrated for the case $n = 7$ in Fig. 10. Circulant graphs having $n$ nodes are defined using modulo $n$ arithmetic, and regular circulant graphs representing grids can be so defined for any odd number $n \geq 3$ [4]. For the case $n = 7$, as illustrated in Fig.10, the nodes are numbered 0 through 6 and arranged clockwise on the circumference of a circle. This numbering makes it possible to use modulo 7 arithmetic to describe whether an edge exists between any two nodes [2]. For odd $n \geq 3$, it is also possible to describe how families of regular circulant graphs with $n$ nodes can be constructed in terms of the number of "hops" between nodes along the circle's circumference. The least densely interconnected regular circulant graph has degree 2, and each node is adjacent only to the 2 nodes reachable by taking one hop clockwise or counterclockwise. Permitting either one or two hops in either direction yields a regular circulant graph of degree 4, and in general, permitting as many as $k$ hops yields a regular circulant graph of degree $2k$. This process can continue until $k = (n-1)/2$, at which point every two nodes are adjacent (in which case the graph is said to be *complete*). In Fig. 10, $(n-1)/2 = 3$, so the family of regular circulant graphs on 7 nodes has the 3 members illustrated.

The stability values of the grids represented by these graphs can be determined explicitly [14] and plotted as a function of both number of nodes $n$ and degree $2k$. Such a plot is shown in Fig. 11 for odd integers $n$, $3 \leq n \leq 19$. For each $n$, stabilities $\alpha_2$ are shown for degrees $2k$, $1 \leq k \leq (n-1)/2$. Notice that for each $n$, stability invariably increases with increasing degree, attaining its maximum value when the grid under consideration is represented by a complete graph. The stabilities of ring grids, i.e., grids represented by regular circulant graphs of degree 2, decrease as the number of nodes in the ring increases. The stabilities of complete grids appear along the diagonal in Fig. 11. The stability value of a complete grid is the same as the number of nodes in the graph. This fact, and the preceding assertions, can be obtained directly from the functional form of the stability values $\alpha_2$. Specifically,

$$\alpha_2 = (2k+1) - \frac{\sin[(2k+1)\pi/n]}{\sin(\pi/n)} \qquad (13)$$

for all permissible values of $n$ and $k$. For a ring grid ($k = 1$), trigonometric identities yield $\alpha_2 = 2 - 2\cos(2\pi/n)$, which explains why stability values of ring grids tend to zero for large $n$. In general, though, stability values depend on a quotient of sine functions. MATLAB provides a "fit" function that can be used to fit various types of familiar surfaces to discrete data plots such as those in Fig. 11. Fig. 12 illustrates the result of fitting a quadratic surface to the data of Fig. 11; the quadratic fit, while not perfect, was better than any exponential fit attempted. This suggests that at least for some regular grids, stability increases approximately quadratically with degree, i.e., with richness of interconnection.

Regular circulant graphs also facilitate the study of how graph theoretic properties other than average degree can affect grid stability. Two such properties, characteristic path length and clustering coefficient, were examined in [2].

## 6. Conclusion

At least one dynamic property of idealized uniform synchronous electric power grids, steady-state stability, can be analyzed using the algebraic connectivity values of their associated graphs. Although these values are difficult to determine for most topologies, analysis of topologies for which they can be determined suggest some interesting possible relationships between idealized grids and real-world grids.

## 7. Acknowledgement

The authors gratefully acknowledge the facilities provided by the Center for Advanced Power Systems at Florida State University (FSU).

# Figures

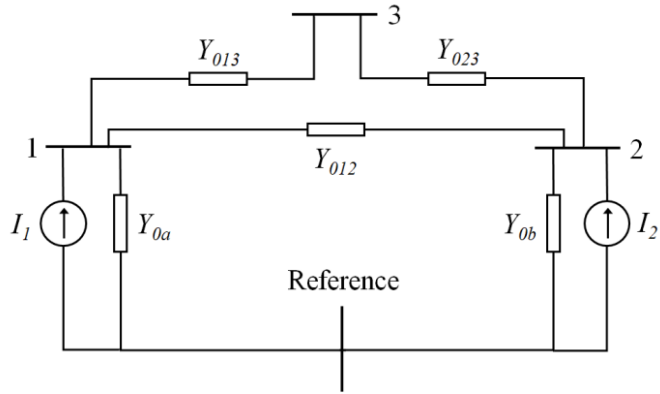

Fig. 1. Example Admittance Diagram

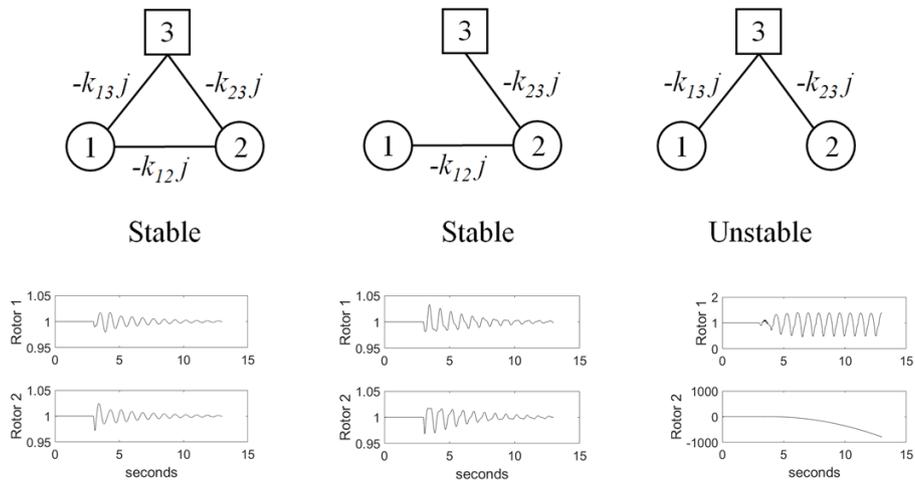

Fig. 2. Stability of Selected Topologies

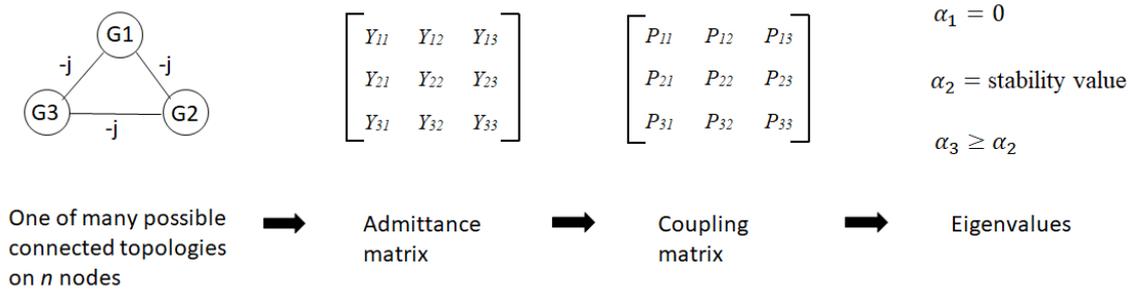

Fig. 3. Mapping grid topologies to stability values $\alpha_2$

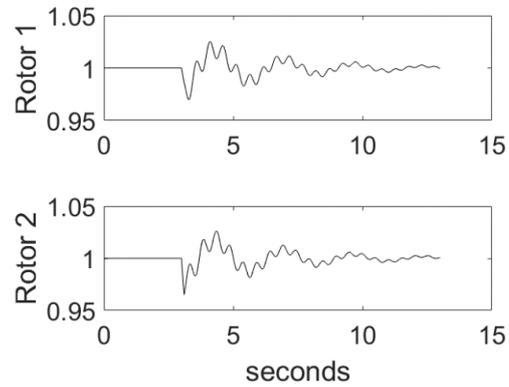

Fig. 4. Rotor angle velocities, $\alpha_2 = 0.0532$

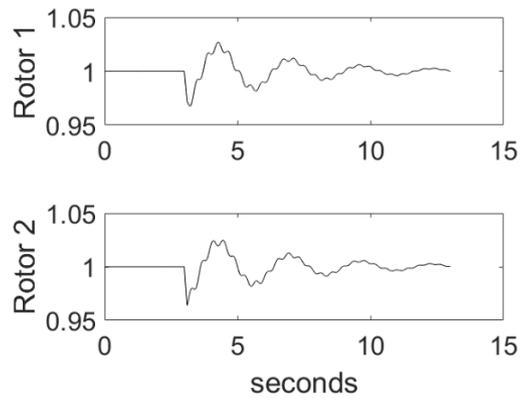

Fig. 5. Rotor angle velocities, $\alpha_2 = 0.264$

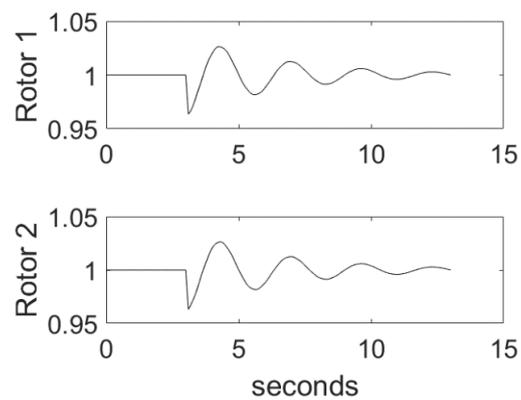

Fig. 6. Rotor angle velocities, $\alpha_2 = 5.3$

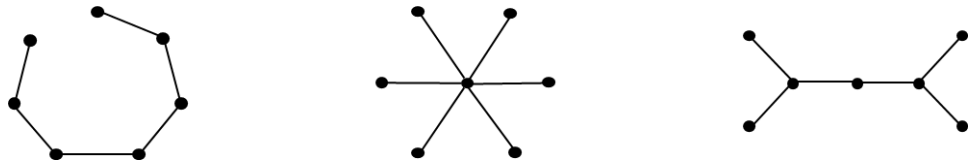

Fig. 7. Some trees having seven nodes

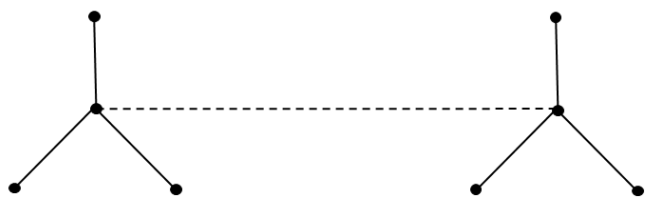

**Grid 1**  **Grid 2**

Fig. 8. Edge that minimizes diameter, maximizes stability

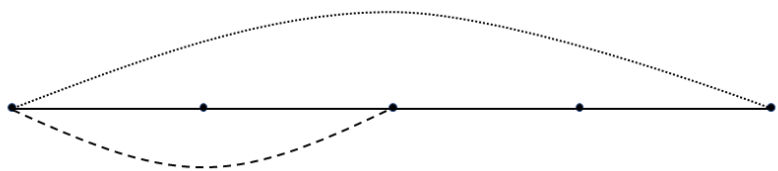

Fig. 9. Stability depends on cycle length

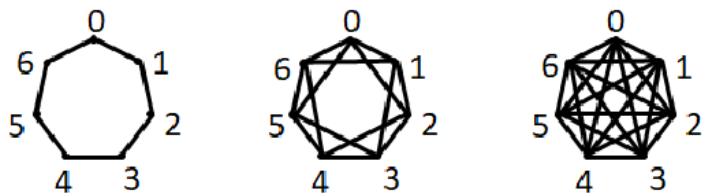

Fig. 10. A family of 7-node regular circulant grids

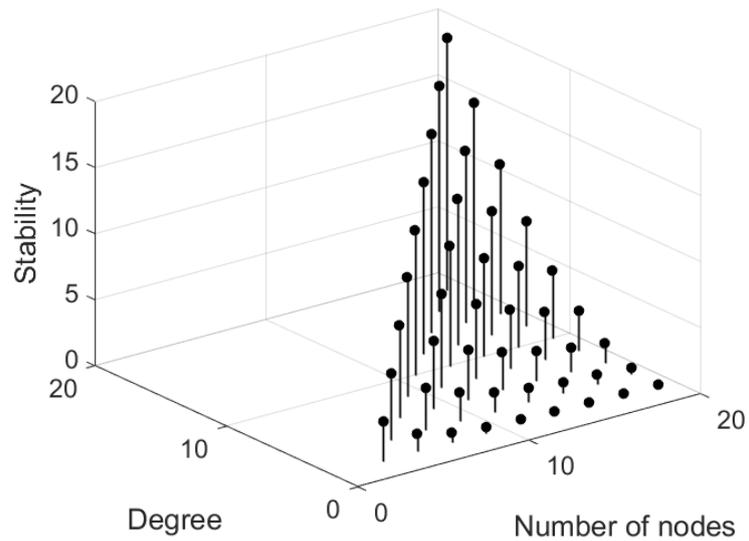

Fig. 11. Stability values of regular circulant grids

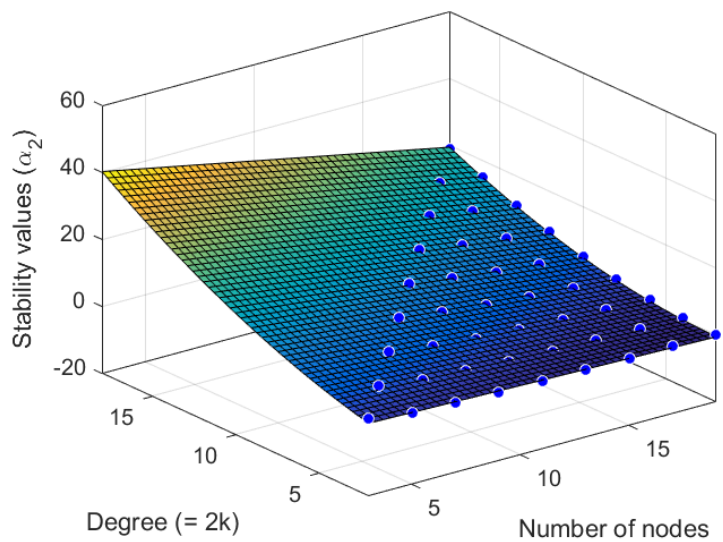

Fig. 12. A quadratic fit of regular circulant grid stability

**BIOGRAPHIES**

**James R. Stright** received a B.E.E. degree from Gannon University in Erie, Pennsylvania and immediately began a career in the U.S. Air Force. While in the Air Force, he obtained a Ph.D. from the Air Force Institute of Technology at Wright-Patterson AFB, Ohio. He conducted and led defense-related research throughout his Air Force career, and continued to do so in the private sector following his Air Force retirement. Much of this research addressed problems related to secure digital communications, real-time image processing, and enhancement of defense-related information processing. He studied power systems engineering at Florida State University with a research emphasis on applications of graph theory to power grid dynamics, receiving a Ph.D. in Electrical Engineering in 2019.

**Chris S. Edrington** received his BS in Engineering from Arkansas State University in 1999 and his MS and PhD in Electrical Engineering from the Missouri University of Science and Technology in 2001 and 2004, respectively, where he was both a DoE GAANN Fellow for his MS focusing on control systems and NSF IGERT Fellow for his PhD focusing on variable speed drive systems. He currently is a Professor of Electrical and Computer Engineering with the FAMU-FSU College of Engineering and is the lead for the Energy Conversion and Integration thrust for the Florida State University-Center for Advanced Power Systems. His research interests include modeling, simulation, and control of electromechanical drive systems; applied power electronics; distributed control; integration of renewable energy, storage, and pulse power loads. Dr. Edrington has published over 170 papers (including 2 IEEE Prize Awards), has graduated 23 MS students and 15 PhD students (with 5 in process) and has 6 patents with an additional 8 patents pending.